\documentclass[sigconf,authorversion]{acmart}

\usepackage[normalem]{ulem}

\AtBeginDocument{%
  \providecommand\BibTeX{{%
    \normalfont B\kern-0.5em{\scshape i\kern-0.25em b}\kern-0.8em\TeX}}}

\copyrightyear{2024} 
\acmYear{2024} 
\setcopyright{acmlicensed}\acmConference[WSDM '24]{Proceedings of the 17th ACM International Conference on Web Search and Data Mining}{March 4--8, 2024}{Merida, Mexico}
\acmBooktitle{Proceedings of the 17th ACM International Conference on Web Search and Data Mining (WSDM '24), March 4--8, 2024, Merida, Mexico}
\acmPrice{15.00}
\acmDOI{10.1145/3616855.3635699}
\acmISBN{979-8-4007-0371-3/24/03}

\usepackage{booktabs}
\usepackage{color}
\usepackage{amsmath}
\usepackage{enumitem}
\usepackage{multirow}

\begin{document}

\title{IAI MovieBot 2.0: An Enhanced Research Platform with Trainable Neural Components and Transparent User Modeling} 

\author{Nolwenn Bernard}
\authornote{Equal contribution.}
\affiliation{%
  \institution{University of Stavanger}
  \city{Stavanger}
  \country{Norway}
}
\email{n.m.bernard@uis.no}

\author{Ivica Kostric}
\authornotemark[1]
\affiliation{%
  \institution{University of Stavanger}
  \city{Stavanger}
  \country{Norway}
}
\email{ivica.kostric@uis.no}

\author{Krisztian Balog}
\affiliation{%
  \institution{University of Stavanger}
  \city{Stavanger}
  \country{Norway}
}
\email{krisztian.balog@uis.no}

\renewcommand{\shortauthors}{Nolwenn Bernard, Ivica Kostric, \& Krisztian Balog}

\begin{abstract}
While interest in conversational recommender systems has been on the rise, operational systems suitable for serving as research platforms for comprehensive studies are currently lacking.
This paper introduces an enhanced version of the IAI MovieBot conversational movie recommender system, aiming to evolve it into a robust and adaptable platform for conducting user-facing experiments.
The key highlights of this enhancement include the addition of trainable neural components for natural language understanding and dialogue policy, transparent and explainable modeling of user preferences, along with improvements in the user interface and research infrastructure.
\end{abstract}



\maketitle

\section{Introduction}

Recent years have seen significant advancements in recommender systems, transitioning from traditional methods that implicitly gauge user preferences through historical data to more interactive, conversational recommender systems (CRSs). These CRSs employ human-machine conversation techniques to elicit explicit user preferences through multi-turn dialogues, providing more tailored recommendations. 
Various aspects of the conversational recommendation problem have been studied, including preference elicitation~\citep{Kostric:2021:RecSys}, multi-turn strategies~\citep{Zhang:2018:CIKM}, natural language understanding and generation~\citep{Gao:2018:SIGIR}, and system evaluation~\citep{Zhang:2020:KDD}.
However, much of the existing research focuses on individual components rather than complete, user-facing systems~\citep{Gao:2021:AI}.
There remain gaps in understanding how the individual parts work together and affect end-to-end performance and user satisfaction.
This, to a large extent, may be attributed to the lack of open source CRSs that can serve as research platforms to enable such studies.

We observe that, to the best of our knowledge, few open-source CRSs are currently available.
Notably, several of these rely on commercial components (e.g., Google Dialogflow and IBM Watson). 
These platforms are often difficult to customize or extend for specific research experiments. Even when it is possible to conduct experiments, the results are often not reproducible due to the inherent black-box nature of these solutions, posing a challenge for scientific validation and further studies. Finally, these commercial services require a paid subscription, which limits their accessibility. 

IAI MovieBot~\citep{Habib:2020:CIKM} is an open-source CRS built on a modular architecture that provides movie recommendations. 
It has been used to perform research on user utterance reformulation~\citep{Zhang:2022:SIGIR} and evaluation of CRSs via user simulation~\citep{Afzali:2023:WSDM}.
Currently, IAI MovieBot relies on rule-based and template-based components, which, while still common, are considered outdated in the rapidly evolving field of natural language processing and conversational AI. 
Moreover, the front-end options are limited to the terminal or the third-party Telegram platform, which requires user registration.

In this paper, we present IAI MovieBot 2.0\footnote{\url{https://iai-group.github.io/MovieBot/}} which includes:
\begin{itemize}[leftmargin=0.5cm]
    \item New natural language understanding and dialogue manager components trained using deep learning approaches, thereby making them more robust to unexpected behavior compared to rule-based components in the previous version of IAI MovieBot. 
    \item A user model to store personal preferences beyond a single conversation. It can be used to streamline the recommendation process, provide explicit control over the preferences stored, and increase transparency, among others. Furthermore, it opens up new research avenues concerning long-term preference management, such as how to resolve conflicts between newly expressed preferences and those already stored.
    \item A new web widget as front-end along with new deployment solutions (REST API and socket.io server) capable of accommodating multimodal interactions and offering extensive customization options, to enable experiments focusing on the interaction between humans and conversational agents.
    \item An updated codebase that utilizes the DialogueKit\footnote{https://pypi.org/project/dialoguekit/} library~\citep{Afzali:2023:WSDM}, making IAI MovieBot 2.0 more modular and easier to modify and extend.
\end{itemize}

\section{Related work}
\label{sec:related}

While there is a growing interest in conversational recommender systems~\citep{Gao:2021:AI}, there is a scarcity of operational systems that can be used as research platforms for comprehensive studies.
Some of the existing CRSs are closed-source commercial products, like And Chill.\footnote{\url{http://www.andchill.io}}
Open-source research prototypes include Vote Goat~\citep{Dalton:2018:SIGIR} and DAGFiNN~\citep{Kostric:2022:RecSys}. 
In terms of objectives, Vote Goat is the most similar to IAI MovieBot.
Both are created for the movie domain and are positioned as research platforms using a modular architecture.
The main difference lies in Vote Goat's use of Google Dialogflow to handle user interactions.
DAGFiNN allows multi-modal interactions, e.g., text and voice, and performs recommendations over multiple domains, specifically, points-of-interest and conference sessions.
The idea of a modular architecture is also present in DAGFiNN as it is built on top of the Rasa framework. %
However, recommendations are just one of the many skills supported and not the sole focus.

There exist a few frameworks that are specifically meant for developing CRSs, such as  
ConvLab-3~\citep{Zhu:2022:arXiv} and  CRSLab~\citep{Zhou:2021:arXiv}. 
Generally, the frameworks provide a set of options for various CRS components. For example, ConvLab-3 includes T5 or BERT-based models for natural language understanding and CRSLab offers five different dialogue policy models.
It is worth noting that the different neural models available are commonly trained on standard datasets; tailoring them to a specific domain requires additional fine-tuning on a suitable dataset. 

\section{The Existing IAI MovieBot}
\label{sec:moviebot}

\begin{figure*}
    \centering
    \includegraphics[scale=0.29]{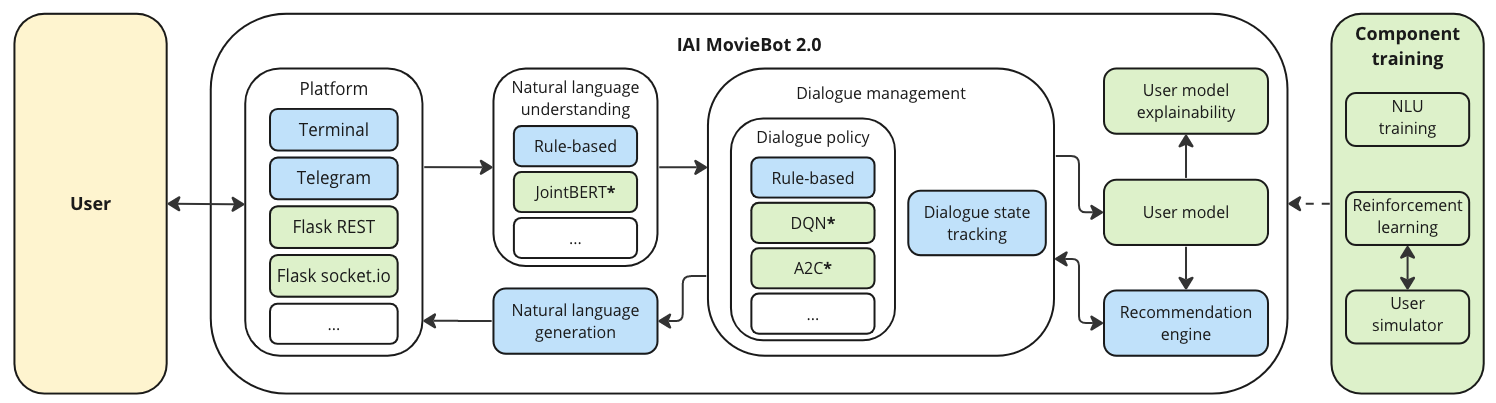}
    \caption{Overview of IAI MovieBot 2.0 architecture. Blue components are inherited from~\citep{Habib:2020:CIKM} and the green ones are new additions. Training utilities are available for components marked with a star (*).}
    \label{fig:moviebot}
    \vspace*{-0.75\baselineskip}
\end{figure*}

IAI MovieBot~\citep{Habib:2020:CIKM} is an open source CRS that aims to provide movie recommendations via multi-turn conversations.
Its architecture consists of four main components (see  Fig.~\ref{fig:moviebot}): natural language understanding (NLU), dialogue manager, recommendation engine, and natural language generation. The NLU uses pattern-based methods to detect intents like  \texttt{REVEAL} or \texttt{INQUIRE}, and identifies values of specific slots such as \texttt{genre}, \texttt{director name}, and \texttt{release year} to understand user preferences. The dialogue manager includes a dialogue state tracker and a rule-based dialogue policy, designed to cater specifically to this task. During the conversation, IAI MovieBot dynamically elicits user preferences and updates the dialogue state to provide relevant movie recommendations,  as user preferences evolve or change.

While IAI MovieBot has been a valuable research platform, it has several limitations that necessitate an updated version.
\begin{itemize}[leftmargin=0.5cm]
    \item \textbf{Rule-/template-based components}: These methods have been surpassed by more modern alternatives that require less manual effort from system designers. Rule-based components require extensive domain knowledge, and can be challenging to update or create complex rules.
    Moreover, template-based components are rigid and thus can hurt the user experience.
    \item \textbf{Limited UI/messaging platform}: The user can interact with IAI MovieBot either via the terminal or Telegram.\footnote{\url{https://t.me/iai_moviebot_bot}} The terminal is not user-friendly, while Telegram is a third-party application that requires a separate registration from the user and has its own constraints in terms of UI that are outside of our control.
    \item \textbf{Limited extensibility}: While MovieBot is designed to be modular, the integration of new methods and technologies is challenging. For example, the NLU component is tightly connected to slot annotations, and some methods are too complex.
\end{itemize}

\section{IAI MovieBot 2.0 Extensions}
\label{sec:extension}

Next, we present the extensions and changes in IAI MovieBot 2.0, which are highlighted in the architecture in Fig.~\ref{fig:moviebot}.
Specifically, we introduce new neural components for natural language understanding and dialogue policy learning (Sections~\ref{sec:nlu} and \ref{sec:DM}), a user model (Section~\ref{sec:um}), as well as improvements to the user interface and  experience (Section~\ref{sec:ux}) and  to research infrastructure (Section~\ref{sec:dx}).

\subsection{Neural Natural Language Understanding}
\label{sec:nlu}

Traditional NLU approaches, including the one used in the initial version of IAI MovieBot, often employ separate models for intent classification and slot-filling. This leads to computational inefficiency and limits contextual understanding capabilities~\citep{Weld:2021:arXiv}. To address these shortcomings, IAI MovieBot 2.0 adopts a unified model that enables simultaneous identification of user intent and extraction of relevant entities or slots.
Specifically, we integrate BERT with a task-specific layer for both intent classification and slot-filling. 
To further refine this model, we incorporate a modified Conditional Random Field (CRF) layer on top of JointBERT~\citep{Chen:2019:arXiv}. The CRF layer is instrumental in capturing the intricate dependencies that often exist between user intents and slots. For example, in the case of a \texttt{REVEAL} intent, where a user explicitly states a preference, the CRF layer ensures that this intent is always associated with at least one annotated slot. This addition not only improves the model's predictive accuracy but also makes the system more robust and enhances its ability to understand context.
Furthermore, performing intent and slot annotations jointly also simplifies the system architecture, making it easier to train, maintain, and extend.

\subsection{Dialogue Policy Learning}
\label{sec:DM}

The dialogue policy is a critical component of a CRS, as it is responsible for selecting the next action based on the current state of the conversation. 
Traditionally, dialogue policies are rule-based, i.e., manually designed by experts (CRS developers), making them tedious to design, maintain, and adapt to new domains.
Therefore, recent work proposed to leverage reinforcement learning (RL) to learn a dialogue policy either from a corpus of historical conversations or interactions with a simulated user~\citep{Sun:2018:SIGIR}.

For IAI MovieBot 2.0, we implement two RL algorithms to learn a dialogue policy: advantage actor critic network (A2C)~\citep{Konda:1999:NIPS} and deep Q-network (DQN)~\citep{Mnih:2015:Nature}.
During the training, we use an episodic reward function based on the success of the conversation; additionally, a reward per turn can be defined.
The episodic reward considers four cases and assigns them a different score: (i) 100, if a recommendation is accepted, (ii) -50, if none of the recommendations are accepted, (iii) -100, if no recommendation are made, and (iv) -1000, if an exception is raised when updating the dialogue state. 
We make a disction between the dialogue state and the Markovian state, that is, the state used by the dialogue policy to select the next action. 
The latter can take the form of either a one-hot encoded vector representing the dialogue state (e.g., a recommendation was made, beginning of the conversation, and should make a recommendation) or a concatenation of the previous vector with two additional one-hot encoded vectors, corresponding to the last user and agent intents.

We perform reinforcement learning using user simulation, as there is no sufficient dataset of historical conversations collected with IAI MovieBot. Thus, we create an agenda-based user simulator on top of the UserSimCRS toolkit~\citep{Afzali:2023:WSDM}.
Additionally, we provide two different environments for training the dialogue policy: one using the NLU component of IAI MovieBot and the other using semantic annotations directly from the user simulator.
The latter allows us to bypass the NLU component and thereby eliminate annotation errors that could potentically impact the update of the dialogue state.
These environments are built on top of Gymnasium~\citep{Towers:2023:git}, which is a well-established library for reinforcement learning.

\subsection{User Modeling}
\label{sec:um}

The original IAI MovieBot does not include an explicit user model; only the preferences expressed in the given session are considered.
However, without a user model, one cannot store long-term preferences and utilize them for future recommendations~\citep{Jannach:2021:CSUR}.
Therefore, we integrate a user model into IAI MovieBot 2.0, inspired by the work of~\citet{Balog:2019:SIGIR}, which is transparent and explainable by design. 
Once authenticated, users have their preferences securely stored for future interactions.
This persistent user model serves as a dynamic repository of user preferences, reducing the need for repetitive preference elicitation in subsequent conversations and allowing for increasingly more personalized recommendations. 
The user model is controled and updated by the dialogue manager via the dialogue state tracker. 
Specifically, we store short-term and long-term preferences of the user both in an unstructured way (raw utterances) and in a structured form (key-value pairs); it is left to the recommendation engine how exactly this information is utilized.
The user model can be displayed to the user (either in a raw form or as a set of summary statements following~\citep{Balog:2019:SIGIR}), thereby providing transparency and control over the recommendation process.

\subsection{User Interface and Experience}
\label{sec:ux}

In the initial version of IAI MovieBot there is a limited range of options when it comes to experiments related to user experience, given that it is served on a third-party messaging platform. 
To tackle this, IAI MovieBot 2.0 comes with an independent web widget that can be added to any website.
This allows for new features and UI elements to be added seamlessly.
For example, a login feature is available to enable the creation of user models.
Transparency and explainability are integral elements of the system's design (cf. Section~\ref{sec:um}) that round out the user experience. 

\subsection{Research Infrastructure Improvements}
\label{sec:dx}

IAI MovieBot is meant to serve as a research platform that can support user-facing experiments on CRSs. Therefore, a main motivation behind the new version is to make the codebase easier to update and extend, to facilitate such experiments.
First, we integrate tools for linting, code formatting, and unit testing to standardize the codebase and to facilitate community contributions more effectively.
Second, the new version uses DialogueKit~\citep{Afzali:2023:WSDM} as its underlying framework, a library focused on base classes for fundamental concepts in conversational information access. DialogueKit's modular architecture allows for greater component management flexibility.
Third, we equipped the system with new deployment solutions: a REST API and a socket.io server.
Finally, the terminal interface has been revamped to facilitate faster development cycles. It is now optimized for quick testing and debugging, allowing for efficient iterations through changes.
This is especially useful for fine-tuning dialogue management and natural language processing modules. Overall, these improvements make IAI MovieBot 2.0 a more modular, extensible, and developer-friendly platform.

\section{Experiments}
\label{sec:eval}

In this section, we report evaluation results for the newly added neural components, namely, JointBERT for intent and slot annotation and dialogue policies learned using reinforcement learning.
Note that our focus is on showcasing the possibility of experimenting with diverse component variants, and not necessarily on improving performance.

\subsection{Natural Language Understanding}

To assess the effectiveness of the unified NLU model, JointBERT, implemented in IAI MovieBot 2.0, we conducted a comparative evaluation against the original rule-based NLU system. The dataset used for this evaluation was collected from the original IAI MovieBot and underwent manual error correction to ensure its quality.

The training data for JointBERT was synthetically generated using ChatGPT. For intents that did not require slot annotations, 30 data points per intent were obtained. For those requiring slot annotations, 30 data points per slot were collected.

Somewhat unexpectedly, we find that the rule-based NLU outperforms JointBERT on every metric; see Table~\ref{tab:nlu-evaluation}. 
However, considering that the rules were crafted specifically for this task and domain, and that the evaluation dataset was collected using that system, the strong performance from the rule-based approach should not come as a surprise. 
On the other hand, JointBERT was trained using a synthetic dataset of limited size. Increasing the training size and quality would likely result in much stronger performance.

Common errors in the rule-based system involve misclassifying the \texttt{REVEAL} slot when detecting the \texttt{keyword} and \texttt{title} slots. This occurs because the database used for lookup contains phrases often used in general conversation, such as the keyword ``sounds like'' or the title ``No Thank You.'' Some errors associated with JointBERT include confusing the \texttt{REVEAL} and \texttt{REMOVE\_PREFERENCES} intents, especially when dealing with negative preferences, or classifying intent correctly but failing to annotate any of the associated slots.

\begin{table}[t]
\centering
\caption{NLU evaluation (intent and slot annotations).}
\label{tab:nlu-evaluation}
\vspace{-0.75\baselineskip}
\resizebox{\columnwidth}{!}{%
\begin{tabular}{ccrrr}
\hline
\textbf{Model} & \textbf{Metric} & \textbf{Precision} & \textbf{Recall} & \textbf{F1-Score} \\
\hline
\multirow{2}{*}{Rule-based NLU} & Intent & 0.818 & 0.808 & 0.813 \\
 & Slot & 0.886 & 0.943 & 0.914 \\
\hline
\multirow{2}{*}{JointBERT} & Intent & 0.556 & 0.550 & 0.553 \\
 & Slot & 0.685 & 0.263 & 0.380 \\
\hline
\end{tabular}
}
\vspace*{-0.75\baselineskip}
\end{table}
\vspace*{-0.5\baselineskip}

\subsection{Dialogue Policy Learning}

To evaluate a dialogue policy trained with reinforcement learning, we use similar metrics as in~\citep{Sun:2018:SIGIR}, specifically, average reward $R$, success rate $S$, and the average number of utterances $U$.
Wrong Quit Rate $W$ is adapted to our problem; it represents the ratio of conversations where the dialogue state tracker fails to be updated correctly.
Table~\ref{tab:eval_rl} presents results for three dialogue policies: A2C, DQN, and a variation of DQN that also includes the last participant intents in the Markovian state.
We observe that while A2C has the lowest $W$, it also has a very low $S$ and $U$.
It may indicate that the policy is terminating the conversation in the first few utterances, hence, the agent does not have the opportunity to make recommendations.
On the other hand, DQN has more successes, but $W$ is also significantly higher.
Finally, we see that adding context to the Markovian state has a positive impact on every metric.
Overall, this experiment shows that the elements comprising the Markovian state strongly influence the ability to learn a good dialogue policy.

\section{Conclusion}
\label{sec:concl}

We have introduced an enhanced version of IAI MovieBot, with the primary goal of evolving it into a robust and adaptable platform for conducting user-facing experiments.
Novel additions include trainable neural components for natural language understanding and dialogue policy, transparent and explainable modeling of user preferences, as well as user interface and research infrastructure improvements.
In future work, we aim to conduct user studies to compare various system configurations (e.g., rule-based vs. neural NLU and different dialogue policies). We also plan to extend our system with additional neural components (e.g., for natural language generation and for item recommendation). \\

\begin{table}[t]
    \caption{Evaluation of dialogue policies trained with RL.}
    \label{tab:eval_rl}
    \vspace{-0.75\baselineskip}
    \centering
    \begin{tabular}{lcrrc}
        \hline
        \textbf{Policy} & \textbf{R} & \multicolumn{1}{c}{\textbf{S}} (\%) & \multicolumn{1}{c}{\textbf{U}} & \textbf{W} (\%) \\
        \hline
        A2C & -180.7 & 1.9 & 3.2 & 18.5 \\
        DQN & -286.3 & 14.4 & 33.3 & 62.8 \\
        DQN\_intents & -88.04 & 50.4 & 17.1 & 30.5 \\
        \hline
    \end{tabular}
\vspace*{-0.65\baselineskip}
\end{table}

\noindent

\begin{acks}
    We thank all contributors of IAI MovieBot.
    This research was partially supported by the Norwegian Research Center for AI Innovation, NorwAI (Research Council of Norway, project number 309834), and by an unrestricted gift from Google.
\end{acks}

\vspace*{-0.7\baselineskip}

\bibliographystyle{ACM-Reference-Format}
\bibliography{wsdm2024-moviebot.bib}

\end{document}